\documentclass[12pt,a4paper]{article}

\title{Superintegrability in a two-dimensional space \\
       of non-constant curvature}

\author{E.~G.~Kalnins\thanks{Email: e.kalnins@waikato.ac.nz},
J.~M.~Kress\thanks{Email: jonathan@math.waikato.ac.nz} \\
{\small Department of Mathematics, University of Waikato,} \\[-4pt]
{\small Private Bag 3105, Hamilton, New Zealand.}
\\ \\
P.~Winternitz\thanks{Email: wintern@crm.umontreal.ca} \\
{\small Centre de recherches math\'ematiques,
Universit\'e de Montr\'eal,} \\[-4pt]
{\small C.P. 6128-CV, Montr\'eal, Qu\'ebec H3C 3J7, Canada.}}

\usepackage{bbm}

\newcounter{commentnumber}
\usepackage{calc}
\usepackage{ifthen}
\newcounter{hour}\newcounter{minute}
\ifthenelse{\theminute<10}
   {}
   {}

\begin{document}

\maketitle

\begin{abstract}
A Hamiltonian with two degrees of freedom is said to be
superintegrable if it admits three functionally independent integrals of
the motion.  This property has been extensively studied in the case of 
two-dimensional spaces of constant (possibly zero) curvature when all the
independent integrals are either quadratic or linear in the canonical
momenta.  In this article the first steps are taken to solve the problem of
superintegrability of this type on an arbitrary curved manifold in two
dimensions.  This is done by examining in detail one of the spaces of
revolution found by G.~Koenigs.  We determine that there are essentially
three distinct potentials which when added to the free Hamiltonian of this
space have this type of superintegrability.  Separation of variables for the
associated Hamilton-Jacobi and Schr\"odinger equations is discussed.  The
classical and quantum quadratic algebras associated with each of these
potentials are determined.
\end{abstract}

\newpage

\section{Introduction}
\label{sec:intro}

A Hamiltonian system in classical mechanics with $n$ degrees of freedom is 
described by a Hamiltonian function 
$H(x_1,\ldots,x_n,p_1,\ldots,p_n)=H(x,p)$.  
The dynamics of such a system is described by Hamilton's equations
\begin{equation}
\dot x_i = \frac{\partial H}{\partial p_i}\,, \qquad
\dot p_i = - \frac{\partial H}{\partial x_i}\,.
\end{equation}
The time rate of change of a classical observable $\ell=\ell(x,p)$ is given 
by
\begin{equation}
\frac{d\ell}{dt} = \{\ell ,H\}
 = \sum^n_{i=1}\left(
       \frac{\partial\ell}{\partial x_i}\frac{\partial H}{\partial p_i}
          - \frac{\partial\ell}{\partial p_i}\frac{\partial H}{\partial x_i}
               \right)
\end{equation}
where $\{\ ,\ \}$ is the Poisson bracket.  A Hamiltonian system is called 
``Liouville integrable'' if it admits $n$ functionally independent 
integrals of motion $\{X_1,\ldots,X_n\}$ which are mutually in 
involution, i.e.
\begin{equation}
\{X_i,X_j\}=0, \quad i,j=1,\ldots,n
\end{equation}
where one of these constants can be taken to be the Hamiltonian $H$
\cite{Goldstein,Arnold}.
The system is
superintegrable if a further $m$ integrals 
$\{Y_1,\ldots,Y_m,1\leq m\leq n-1\}$ 
exist such that the set of constants $\{X_1=H,X_2,\ldots,X_n,Y_1,\ldots,Y_m\}$ 
are functionally independent.
The additional integrals have vanishing Poisson bracket 
with $H$, but not necessarily with each other or with the $X_i$'s.  A 
classical Hamiltonian system is maximally superintegrable if $m=n-1$.
There are 
then $2n-1$ functionally independent integrals of motion.
The concepts of complete
integrability and superintegrability have their analogue in quantum 
mechanics.  
In this case a superintegrable quantum mechanical system is described
by $n+m$ quantum observables 
$\{\hat X_1=\hat H,\hat X_2,\ldots,\hat X_n,\hat Y_1,\ldots,\hat Y_m\}$ 
which satisfy the commutation relations
\begin{equation}
[\hat H,\hat X_i]=\hat H\hat X_i-\hat X_i\hat H=0\,, \quad
[\hat H,\hat Y_j]=0\,, \quad
[\hat X_i,\hat X_k]=0
\end{equation}
where $i,k=1,..,n,j=1,..,m$.  For superintegrable classical 
Hamiltonian systems it is often the case that the elements of our set of 
constants are polynomial in the canonical momenta.  The best known maximally 
superintegrable systems in Euclidean space $E_n$ are the Kepler problem and the
harmonic oscillator.  
All finite (bounded) trajectories in these two systems are 
closed.
Moreover these are the only spherically symmetric potentials for which 
all finite trajectories are closed \cite{Bertrand}. 

Systematic studies of superintegrable systems have been conducted for spaces 
of constant curvature in two and three dimensions 
\cite{FMSUW,WSUF,MSVW,Evans90,Evans91,KMPW99,KMP96}.  
In particular, a complete classification all superintegrable
systems in the real Euclidean spaces $E_2$ and $E_3$ with at most second order
integrals of motion was given 
\cite{FMSUW,WSUF,MSVW,Evans90,Evans91}.
More recently a 
relation between superintegrable systems and generalized Lie symmetries has 
been established \cite{STW01}, as well as their relation to exactly solvable problems in 
quantum mechanics \cite{TTW}.
Recently \cite{KMP00a,KMP00b,KKMP01} it has been possible to classify all 
maximally superintegrable systems for spaces of constant curvature (possibly 
zero) in two dimensions for which all the extra constants of the motion are at 
most quadratic in the canonical momenta.  

A natural question to ask is whether 
the concept of superintegrability is restricted to spaces of constant 
curvature.  The purpose of this article is to show that this is not so and to 
start a study of superintegrable systems in more general Riemannian, 
pseudo-Riemannian and complex Riemannian spaces.
More specifically, we consider real two-dimensional 
spaces and search for Hamiltonian systems allowing additional 
constants of the motion that are at most quadratic in the momenta.  

To make 
initial progress on this problem we first need to know which Riemannian spaces 
in two dimensions have associated with them more than one 
classical quadratic constant of the motion.  This is a problem that has been comprehensively solved by 
Koenigs \cite{Koenigs} in a note written in volume IV of the treatise of 
Darboux \cite{Darboux}.  

In 
addition to being of intrinsic interest, additional motivation for this problem 
comes from the observation that all two-dimensional Riemannian spaces can be 
embedded in the three-dimensional Euclidean or pseudo-Euclidean space.  
Consequently any such two-dimensional classical motion is equivalent to a
constrained motion in three dimensions.
It is also possible to interpret the 
motion, via general relativity, as motion in a two-dimensional gravitational 
field.

Given that we have a Riemannian space in two dimensions with infinitesimal
distance
\begin{equation}
ds^2=g_{ij}(u)du^idu^j, \quad i,j=1,2
\end{equation}
and $u=(u^1,u^2)$, the classical Hamiltonian has the form 
\begin{equation}
H=\frac12 g^{ij}p_ip_j+V(u)
\end{equation}
and the corresponding Schr\"odinger equation can be taken to have the form
\begin{equation}
\hat H\Psi = -\frac1{2\sqrt g}\partial_{u^i}
  \left(\sqrt g g^{ik}\partial_{u^k}\Psi \right) + V(u)\Psi = E\Psi
\end{equation}
where $g=\det (g_{ij})$.  For the classical Hamiltonian $H$ our 
problem is to look for potentials $V(u)$ and Riemannian spaces specified by the
metric $g_{ij}$ for which there are at least two extra functionally independent
constants of the motion of the form
\begin{equation}
\lambda_1=a^{ij}(u)p_ip_j+b(u)
\end{equation}
or
\begin{equation}
\lambda _2=a^i(u)p_i+c(u)\,,
\end{equation}
other than $H$.
One well known way of solving the corresponding classical problem is to use 
Hamilton-Jacobi theory.  The crucial equation to solve is then the 
Hamilton-Jacobi equation obtained from the equation $H=E$ via the 
substitution $p_i= \frac{\partial S}{\partial u^i}$, that is
\begin{equation}
H=\frac12g^{ij}\frac{\partial S}{\partial u^i}
                 \frac{\partial S}{\partial u^j} + V(u) = E\,.
\end{equation}
This equation is sometimes solvable by the method of separation of variables 
using the additive separation Ansatz 
\begin{equation}
S=S_1(u^1,\alpha ,E)+S_2(u^2,\alpha ,E)\,.
\end{equation}
The corresponding Schr\"odinger 
equation can also be solved by separation of variables with the product 
Ansatz
\begin{equation}
\Psi = \psi_1(u^1,\lambda,E)\psi_2(u^2,\lambda,E)\,.
\end{equation}
The quantities $\lambda_i$ are constants of the motion if
\begin{equation}
\{\lambda _i,H\}=0\,.
\end{equation}
For $\lambda_2$ this implies that $a^i(u)$ is a Killing vector and 
$a^i(u)p_i$ is a symmetry of the free Hamiltonian ($H$ without $V(u)$). 
In the 
case of $\lambda _1$ this implies that $a^{ij}(u)$ is a Killing tensor.
Such
tensors are directly related to the notion of additive separation as described 
above.  We note that for constants of the type $\lambda _2$, the condition 
implies $c(u)=0$.  It is also clear that for every constant linear in the momenta,
its square is a constant quadratic in the momenta, that is, of the form of $\lambda _1$.  

As mentioned 
earlier, Darboux and Koenigs have given a comprehensive analysis of when a two-dimensional 
Riemannian space admits more than one quadratic constant.  
In Section \ref{sec:geodesics} we 
summarise some of these results \cite{Koenigs,Darboux}.
In the remaining sections we
concentrate on a particular space with a Killing vector and two Killing 
tensors.  Section \ref{sec:free} deals with the 
free Hamilton-Jacobi equation and we show that the 
Schr\"odinger equation allows separation of variables in three different 
coordinates systems which we determine explicitly.  Potentials that allow 
separation of variables in these systems are then introduced.
In Section \ref{sec:superint} we find all potentials with this
superintegrability property.
We then
discuss in Section \ref{sec:embed} 
the various surfaces that may be represented by the infinitesimal 
distances that we have and the consequent special functions that arise from the
corresponding Schr\"odinger equation.

\section{On geodesics with quadratic integrals}
\label{sec:geodesics}

In 1889 G.~Koenigs \cite{Koenigs} 
wrote a note in the last volume of Darboux's treatise 
``Th\'eorie g\'en\'erale de surfaces'', the title of which
coincides with the title of this section.  
This note contains a summary of results which are the solution of the 
problem outlined in the Introduction, viz.\ 
when does the free Hamiltonian of a 
two-dimensional Riemannian space admit more than one quadratic constant of the 
motion.  
The analysis was performed over the field of complex numbers and
must be modified over the reals.
What Koenigs did was to write the infinitesimal distance 
for a general two-dimensional Riemannian space in the form
\begin{equation}
ds^2=4f(x,y)dxdy\,.
\end{equation}
This can always be done in two dimensions over $\mathbbm C$.
The corresponding free Hamiltonian then has the form
\begin{equation}
H=\frac1{2f(x,y)}\ p_xp_y\,.
\end{equation}
By making the requirement that there is a second order Killing
tensor of the form 
\begin{equation}
\lambda =a^{ij}(u)p_ip_j\,,
\end{equation}
Darboux and Koenigs establish the following propositions.
\begin{enumerate}
\item
Any two-dimensional Riemannian space that admits more than one Killing 
vector must be a space of constant curvature and admit three linearly 
independent Killing vectors.
\item
Any two-dimensional Riemannian space that admits more than three Killing 
tensors is a space of constant curvature.  It then actually admits five linearly 
independent Killing tensors which are all bilinear expressions in the Killing 
vectors.  The sixth bilinear combination is the Hamiltonian itself.
\item
Any two-dimensional Riemannian space that admits precisely three linearly 
independent Killing tensors will be a Riemannian space of revolution.  In fact 
there will be one Killing vector and two Killing tensors.
\end{enumerate}

Two-dimensional Riemannian spaces of this latter type were distinguished
to be of four types. 
The infinitesimal distances of these types are given by

\begin{itemize}
\item[{(I)}] $\displaystyle ds^2=(x+y)dxdy$.
\item[{(II)}] $\displaystyle ds^2=\left(\frac a{(x-y)^2} + b\right)dxdy$.
\item[{(III)}] $\displaystyle ds^2
     = \left(ae^{-\frac{x+y}2}+be^{-x-y}\right)dxdy$.
\item[{(IV)}] $\displaystyle ds^2
     = \frac{a\left(e^{\frac{x-y}2}+e^{\frac{y-x}2}\right) + b}
            {\left(e^{\frac{x-y}2}-e^{\frac{y-x}2}\right)^2}\ dxdy$.
\end{itemize}
It is the first of these infinitesimal distances that we analyse in some detail 
in the next section.  We shall call the spaces ``Darboux spaces'' and denote then by
$D_1$, $D_2$, $D_3$ and $D_4$ respectively. 

\section{The free particle and separation of variables in a Darboux space of type one}
\label{sec:free}

If we consider the first space of Darboux's 
list and look at real forms of this space only,
it is convenient to make the new choice of variables 
\begin{equation}
x=u+iv, \quad y=u-iv\,.
\end{equation}
The corresponding infinitesimal distance can then be taken as 
\begin{equation}
ds^2=2u(du^2+dv^2)\,,
\end{equation}
and the corresponding Hamiltonian has the form
\begin{equation}
H=\frac1{4u}(p^2_u+p^2_v)\,.
\end{equation}
Associated with this Hamiltonian are three integrals of the free motion, two 
quadratic and one linear.
\begin{eqnarray}
K &=& p_v  \nonumber \\
X_1 &=& p_up_v - \frac v{2u}(p^2_u+p^2_v)  \nonumber \\
X_2 &=& p_v(vp_u-up_v) - \frac{v^2}{4u}(p^2_u+p^2_v)\,.
\end{eqnarray}
These three integrals satisfy the polynomial Poisson algebra 
relations,
\begin{equation}
\{K,X_1\}=2H\,, \quad \{K,X_2\}=-X_1\,, \quad \{X_1,X_2\}=2K^3\,.
\end{equation}
They cannot be functionally independent and in fact satisfy the 
relation
\begin{equation}
4HX_2+X^2_1+K^4=0\,.
\end{equation}
For the analogous quantum problem it is sufficient to consider the operators
\begin{eqnarray}
\hat H &=& -\frac1{4u}\left(\partial ^2_u+\partial ^2_v\right)  \nonumber \\
\hat K &=& -i\partial_v  \nonumber \\  
\hat X_1 &=& -\partial _u\partial _v
           + \frac v{2u}\left(\partial ^2_u+\partial ^2_v\right)  \nonumber \\
\hat X_2 &=& -\frac12\left[\partial _v,v\partial _u-u\partial _v\right]_+
       + \frac{v^2}{4u}(\partial ^2_u+\partial ^2_v)
\end{eqnarray}
where $[A,B]_+=AB+BA$.  The quantum versions of the quadratic constants are 
obtained via the formula
\begin{equation}
\hat \lambda = -\frac1{\sqrt g}
       \partial _i\left(a^{ij}\sqrt{g} \partial _j\right)\,.
\end{equation}
These operators have the same commutation relations as for the classical 
constants with the Poisson bracket replaced by 
the commutator bracket.
\begin{equation}
[\hat K,\hat X_1] = 2i\hat H\,, \quad
[\hat K,\hat X_2] = -i\hat X_1\,, \quad
[\hat X_1,\hat X_2] = -2i\hat K^3\,.
\end{equation}
There is also the operator relation
\begin{equation}
4\hat H\hat X_2+\hat X^2_1+\hat K^4=0\,.
\end{equation}
The question we address in this section relates to the various possible ways that
separation of variables can be achieved in the case of free classical motion or
its quantum analogue, the free Schr\"odinger equation.
The criteria for this to 
occur is the same in either case.
Classically, if we have a general quadratic 
first integral $\lambda $ and free Hamiltonian 
\begin{equation}
H=\frac12 g^{ij}(u)p_ip_j\,,
\end{equation}
and if the characteristic equation,
\begin{equation}
\left|a^{ij}-\rho g^{ij}\right|=0\,,
\end{equation}
has two distinct roots $\rho_1$ and $\rho_2$, the Hamiltonian
will have Liouville form when written in terms of the new variables
$\rho_1,\rho_2$.  That is
\begin{equation}
H = \frac{\sigma(\rho_1)p^2_{\rho_1}+\tau(\rho_2)p^2_{\rho_2}}
         {\rho _1+\rho _2}\,.
\end{equation}
In this form, both classical and quantum systems can be solved by the 
separation of variables Ansatz.

If we want to classify all different separable coordinate systems
for a given Hamiltonian
we need to know how many essentially different quadratic first 
integrals are possible.
To decide on the notion of equivalence we first observe 
that the variable $v$ does not explicitly appear in the metric tensor,
that is, it is an ignorable variable.
This means that the transformations $v\rightarrow v+b$
form a one-dimensional Lie group.
Accordingly, we determine the notion of 
equivalence to mean that two quadratic integrals are equivalent if they are 
related by a motion of this group.  Consequently the most general quadratic 
constant can be written 
\begin{equation}
\lambda =aX_1+bX_2+cK^2
\end{equation}
to within the addition of a multiple of $H$.
The second order elements $X_i$ 
transform under the adjoint action according to
\begin{eqnarray}
X_i &\rightarrow& \exp(\alpha K)X_i\exp(-\alpha K)  \nonumber \\
    &=& \exp (\alpha Ad(K))X_i  \nonumber \\
    &=& X_i+\alpha \{K,X_i\}+\frac12\alpha ^2\{K,\{K,X_i\}\}+\cdots
\end{eqnarray}
or specifically
\begin{eqnarray}
X_1 &\rightarrow& X_1+2\alpha H  \nonumber \\  
X_2 &\rightarrow& X_2-\alpha X_1-\alpha ^2H\,.
\end{eqnarray}
There are three classes of possible quadratic first integrals under this 
equivalence relation.  Typical representatives are
\begin{equation}
\label{reps}
X_1+aK^2\,, \qquad
X_2+aK^2\,, \quad
K^2\,.
\end{equation}
We can now explicitly demonstrate the separable coordinates in each of these 
cases.

\begin{enumerate}
\item
{\em Separating coordinates associated with $X_1+aK^2$}

If we choose a representative to be
\begin{equation}
L=X_1+\sinh c\ K^2\,,
\end{equation} 
the corresponding roots of the characteristic equation and 
hence new variables are
\begin{eqnarray}
r = \rho _1 &=& -2(Cu+v)  \nonumber \\
s = \rho _2 &=& \frac2C(u-Cv)\,, \qquad C=e^{-c}\,.
\end{eqnarray}
In terms of these coordinates the Hamiltonian has the form
\begin{equation}
H = \frac{2(C^2+1)^2}{C(s-r)}\left(\frac1{C^2}p^2_s+p^2_r\right)
\end{equation}
and the corresponding quadratic constant in these coordinates is
\begin{equation}
L = 2 \frac{(C^2+1)^2}{C(s-r)}\left(\frac r{C^2}p^2_s+sp^2_r\right)\,.
\end{equation}
\item
{\em Separating coordinates associated with $X_2+aK^2$}

Taking the second representative in the list (\ref{reps}), that is
$L=X_2+aK^2$, a convenient choice of new variables $\xi ,\eta $ are related to
the roots $\rho _i$ by
\begin{equation}
\rho_1 = \eta^2(2a-\eta^2)\,, \qquad \rho_2=-\xi^2(2a+\xi^2)\,.
\end{equation}
The corresponding classical Hamiltonian then has the form
\begin{equation}
H = \frac{p^2_\xi +p^2_\eta}{2(\xi ^2+\eta ^2)(\xi ^2-\eta ^2+2a)}\,.
\end{equation}
The associated constant of the motion in the new coordinates $\xi $ and 
$\eta$ is
\begin{equation}
L=\frac{\eta ^2(2a-\eta ^2)p^2_\xi -\xi ^2(\xi ^2+2a)p^2_\eta}
       {2(\xi ^2+\eta ^2)(\xi ^2-\eta ^2+2a)}\,.
\end{equation}
The defining coordinates $u,v$ are written in terms of the new coordinates 
$\xi ,\eta $ via 
\begin{equation}
u=\frac12(\xi ^2-\eta ^2)+a\,, \quad v=\xi \eta
\end{equation}
which looks like displaced parabolic coordinates in the $u,v$ plane.
\item
{\em Separating coordinates associated with $K^2$}

For the last representative, $K^2$, we need only the coordinates $u,v$ and to 
recognise the fact that $K=p_v$.  
\end{enumerate}

We conclude this section by discussing the 
solutions to the free particle and free Schr\"odinger equation in these three 
cases.

In case 1 above it is more convenient to choose the variables according to 
\begin{equation}
u=r\cos\theta +s\sin\theta\,, \quad v=-r\sin\theta +s\cos\theta\,.
\end{equation}
The classical Hamilton-Jacobi equation then has the form
\begin{equation}
H=\frac{\left(\frac{\partial S}{\partial r}\right)^2
          + \left(\frac{\partial S}{\partial s}\right)^2}
       {4(r\cos\theta +s\sin\theta)} 
 = E
\end{equation}
which has the general separable solution 
\begin{equation}
S=S_1(r)+S_2(s)= 
     \frac{(4Er\cos\theta-\lambda)^{3/2}}
          {6E\cos\theta}
   + \frac{(4Es\sin\theta+\lambda)^{3/2}}
          {6E\sin\theta}\,.
\end{equation}
The corresponding free Schr\"odinger equation
\begin{equation}
\hat H\Psi =-\frac1{4(r\cos\theta +s\sin\theta)} 
(\partial ^2_r+\partial ^2_s)\Psi =E\Psi
\end{equation}
has the typical product solutions
$$
\Psi = \sqrt{\left(r-\frac\mu{4E\cos\theta}\right)
             \left(s+\frac\mu{4E\sin\theta}\right)}\
  C_{\frac13}\left(
   \frac23\sqrt{4E\cos\theta}\left(r-\frac\mu{4E\cos\theta}\right)^{3/2}
                    \right)
$$
\begin{equation}
\times C_{\frac13}\left(\frac23\sqrt{4E\sin\theta}
              \left(s+\frac\mu{4E\sin\theta}\right)^{3/2}\right)\,,
\end{equation}
where $C_\nu(z)$ is a solution of Bessel's equation.

In the second case the classical Hamilton-Jacobi equation is
\begin{equation}
H = \frac{\left(\frac{\partial S}{\partial\xi}\right)^2
           + \left(\frac{\partial S}{\partial\eta}\right)^2}
         {2(\xi ^2+\eta ^2)(\xi ^2-\eta ^2+2c)} = E
\end{equation}
and has a general solution of the form
\begin{equation}
S = \int \sqrt{2E\xi^4+2Ec\xi^2-\lambda }\ d\xi
   + \int \sqrt{-2E\eta^4+2Ec\eta^2+\lambda}\ d\eta\,,
\end{equation}
which can be expressed in terms of elliptic integrals.
The corresponding 
Schr\"odinger equation has a solution of the form 
$\Psi =\psi _1(\xi )\psi _2(\eta )$ where the $\psi _i$ satisfy
\begin{eqnarray}
(\partial ^2_\xi +2E\xi ^4+4Ec\xi ^2+\lambda )\psi _1(\xi ) &=& 0  \nonumber \\
(\partial ^2_\eta -2E\eta ^4+4Ec\eta ^2-\lambda )\psi _2(\eta ) &=& 0\,.
\end{eqnarray}
These equations are readily identified as the equations for the anharmonic 
oscillator.  

In the third case the classical Hamilton-Jacobi equation is
\begin{equation}
H = \frac1{4u}\left(\left(\frac{\partial S}{\partial u}\right)^2
                    +\left(\frac{\partial S}{\partial v}\right)^2\right) = E
\end{equation}
which has separable solutions 
\begin{equation}
S=\frac1{6E}(4Eu-k^2)^{3/2}+kv\,.
\end{equation}
The separable solutions to the corresponding free Schr\"odinger equation
\begin{equation}
-\frac1{4u}(\partial ^2_u+\partial ^2_v)\Psi = E\Psi
\end{equation}
have the form
\begin{equation}
\Psi =\sqrt{u-\frac{m^2}{4E}}\
          C_{\frac13}\left(\frac23\sqrt{4E}
              \left(u-\frac{m^2}{4E}\right)^{3/2}\right)e^{imv}\,.
\end{equation}

It is clear that the actual solutions to the classical motion or the 
corresponding Schr\"odinger equation depend on the range of values 
assumed by the
various real variables, that is, on
exactly which real manifold we are considering.

\section{Integrable and superintegrability systems for the Darboux space of type one}
\label{sec:superint}

In this section we address the problem of superintegrability for the 
Hamiltonian
\begin{equation}
H=\frac1{4u}(p^2_u+p^2_v)\,,
\end{equation}
that is, look for potentials $V(u,v)$ for which 
\begin{equation}
\bar H=H+V(u,v)
\end{equation}
admits at least two extra quadratic integrals.
The way to solve this problem is 
as follows.
First we consider that we already have one quadratic first integral
\begin{equation}
\label{Lbar}
\bar L=a(u,v)p^2_u+b(u,v)p_up_v+c(u,v)p^2_v+d(u,v)\,.
\end{equation}
We know that the quadratic part of $\bar L$ (i.e.\ that part obtained
by putting $d(u,v)=0$ in (\ref{Lbar})) 
must correspond to 
one of the three possibilities outlined in the previous section.
For each of 
these possibilities separation of variables is possible in coordinates 
$\alpha ,\beta $ where $u=u(\alpha ,\beta )$, $v=v(\alpha ,\beta )$.
The addition 
of a potential implies that separation is preserved.
As a consequence of this 
$\bar H$ can be written as
\begin{equation}
\label{Hbar}
\bar H = \frac{p^2_\alpha +p^2_\beta +f(\alpha )+g(\beta )}
              {\sigma (\alpha )+\tau (\beta)}
\end{equation}
and the corresponding first integral will be
\begin{equation}
\bar L = \frac{\sigma (\alpha )\Bigl(p^2_\beta +g(\beta )\Bigr)
                   -\tau (\beta )\Bigl(p^2_\alpha +f(\alpha )\Bigr)}
              {\sigma (\alpha )+\tau (\beta )}\,.
\end{equation}
The next step is to impose the condition that there is a further quadratic 
first integral and see what conditions this imposes on the functions 
$f(\alpha )$ and $g(\beta )$.  If we do these calculations systematically we 
arrive at the following three cases.
\begin{enumerate}
\item
\begin{equation}
\label{potential1}
H = \frac{p^2_u+p^2_v}{4u}  + \frac{b_1(4u^2+v^2)}{4u} + \frac{b_2}u
   + \frac{b_3}{uv^2}\,.
\end{equation}
The additional constants of the motion have the form
\begin{eqnarray}
R_1 &=& X_2 - \frac{b_1v^4}{4u} - \frac{b_2v^2}u - \frac{b_3(4u^2+v^2)}{v^2u}
           \nonumber \\
R_2 &=& K^2 + b_1v^2 + \frac{4b_3}{v^2}
\end{eqnarray}
and the corresponding quadratic algebra \cite{Dask01,LV95} 
relations are determined by
\begin{eqnarray}
\{R,R_1\} &=& 8HR_1+6R^2_2+16b_2R_2-32b_1b_3  \nonumber \\
\{R,R_2\} &=& -8HR_2-16b_1R_1  \nonumber \\
R^2 &=& -16HR_1R_2-4R^3_2-16b_2R^2_2-64b_3H^2-16b_1R^2_1 \nonumber \\
    & & \quad +64b_1b_3R_2+256b_1b_2b_3 
\end{eqnarray}
where $R=\{R_1,R_2\}$.  The Hamiltonian clearly separates in the coordinates $u$ 
and $v$ as well as the coordinates $\xi,\eta$ given by
$u=\frac12(\xi ^2-\eta ^2)+a$, $v=\xi \eta$. 
This can be seen from the explicit form
\begin{equation}
H = \frac{p^2_\xi+p^2_\eta}
         {2(\xi ^2+\eta ^2)(\xi ^2-\eta ^2+2a)}
   + \frac{b_1\Bigl( (\xi^2-\eta^2+2a)^2+\xi^2\eta^2 \Bigr)
              + 4b_2 + \frac{4b_3}{\xi^2\eta^2}}
          {2(\xi ^2-\eta ^2+2a)}\,.
\end{equation}
The corresponding quadratic quantum algebra relations are
\begin{eqnarray}
[\hat R,\hat R_1] &=& -6\hat R^2_2 - 8\hat H\hat R_1 + 16b_2\hat R_2
             + 2b_1(3+16b_3)  \nonumber \\ {}
[\hat R,\hat R_2] &=& 8\hat H\hat R_2 - 16b_1\hat R_1  \nonumber \\
\hat R^2 &=&  + 4\hat R^3_2 - 8\hat H\left[\hat R_1,\hat R_2\right]_+ - 16b_2\hat R^2_2
             - 16b_1\hat R^2_1 \\
    & & \  - 4b_1(11+16b_3)\hat R_2 - 4(3+16b_3)\hat H^2 + 16b_1b_2(3+16b_3)
   \nonumber
\end{eqnarray}
where $\hat R=[\hat R_1,\hat R_2]$.
\item
\begin{equation}
\label{potential2}
H = \frac{p^2_u+p^2_v}{4u} +  \frac{a_1}u + \frac{a_2v}u
       + \frac{a_3(u^2+v^2)}u\,.
\end{equation}
The additional constants of the motion have the form
\begin{eqnarray}
R_1 &=& X_1- \frac{2a_1v}u + \frac{2a_2(u^2-v^2)}u
         + \frac{2a_3v(u^2-v^2)}u  \nonumber \\
R_2 &=& K^2+4a_2v+4a_3v^2 
\end{eqnarray}
and the corresponding quadratic algebra relations are determined by
\begin{eqnarray}
\{R,R_1\} &=& -8H^2+16a_3R_2+8(a^2_2+4a_1a_3)  \nonumber \\
\{R,R_2\} &=& 16a_2H-16a_3R_1  \nonumber \\
R^2 &=& 16H^2R_2-16a_3R^2_2+32a_2HR_1-16a_3R^2_1  \nonumber \\
 & & \quad -16(a^2_2+4a_1a_3)R_2-64a_1a^2_2
\end{eqnarray}
If we change the coordinates according to $u=r\cos\theta +s\sin\theta$,
$v=-r\sin\theta +s\cos\theta$ the Hamiltonian assumes the form
\begin{equation}
H = \frac{p^2_r +p^2_s + 4a_1 + 4a_2(-r\sin\theta +s\cos\theta )
             + 4a_3(r^2+s^2)}
         {4(r\cos\theta +s\sin\theta)}
\end{equation}
which clearly also separates in these coordinates.

The commutation relations of the corresponding quantum algebra are
\begin{eqnarray}
[\hat R,\hat R_1] &=& 16a_3\hat R_2 + 8\hat H^2 - 8(a^2_2+4a_1a_3)  \nonumber \\ {}
[\hat R,\hat R_2] &=& - 16a_3\hat R_1 + 16a_2\hat H  \nonumber \\
\hat R^2 &=& - 16a_3\hat R^2_2  - 16a_3\hat R^2_1 + 16\hat H^2\hat R_2
               + 32a_2\hat H\hat R_1  \nonumber \\
 & & \quad - 16(a^2_2+4a_1a_3)\hat R_2 + 64(a^2_3-a_1a^2_2)\,.
\end{eqnarray}
\item
The third potential gives rise to a Hamiltonian of the form
\begin{equation}
H = \frac{p^2_u+p^2_v}{4u} + \frac au\,.
\end{equation}
There are three extra constants associated with this Hamilonian,
\begin{equation}
R_1 = X_1 - \frac{2av}u\,, \quad R_2 = X_2 - \frac{av^2}u \quad \mbox{and} \quad K\,.
\end{equation}
The associated Poisson bracket relations are
\begin{eqnarray}
\{K,R_1\} &=& 2H  \nonumber \\
\{K,R_2\} &=& -R_1  \nonumber \\
\{R_1,R_2\} &=& 2K(K^2+2a)
\end{eqnarray}
and the corresponding functional relation amongst these constants is
\begin{equation}
4HR_2+R^2_1+K^4+4aK^2=0\,.
\end{equation}
The commutation relations associated with the corresponding quantum problem 
have the form
\begin{eqnarray}
[\hat K,\hat R_1] &=& 2i\hat H  \nonumber \\ {}
[\hat K,\hat R_2] &=& -i\hat R_1  \nonumber \\ {}
[\hat R_1,\hat R_2] &=& -2i\hat K(\hat K^2-2a)
\end{eqnarray}
and the identity amongst the defining operators is
\begin{equation}
4\hat H\hat R_2+\hat R^2_1+\hat K^4-4a\hat K^2=0\,.
\end{equation}
\end{enumerate}

Upon examination of the various superintegrable potentials we have
constructed we see that by multiplying the equation $H=E$ by a
suitable factor we essentially recover a variant of one of the
superintegrable systems already classified for spaces of constant
(or zero) curvature.  For the first potential above, 
the equation $H=E$ may be written
\begin{equation}
p_u^2+p_v^2+b_1(4u^2+v^2)+4b_2+\frac{4b_3}{v^2}-4Eu=0\,.
\end{equation}
This equation is known to have separable solutions in coordinates
$u$, $v$ and associated parabolic coordinates $\xi,\eta$ given
by $u=\frac12(\xi^2-\eta^2)$, $v=\xi\eta$.  With the second potential,
$H=E$ becomes
\begin{equation}
p_u^2+p_v^2+4a_3(u^2+v^2)+4a_1+4a_2v-4Eu=0
\end{equation}
and the third,
\begin{equation}
p_u^2+p_v^2-4Eu+4a=0\,.
\end{equation}
This observation is crucial to the whole programme
that we will undertake which aims at finding all superintegrable
systems associated with a curved space in two dimensions and having
quadratic constants.

All three of the above systems are special cases of the superintegrable
systems found in $E_2$ \cite{FMSUW,STW01}. They were shown to be exactly solvable in Ref.
\cite{TTW}.

\section{Embeddings of a Darboux space of revolution of type one}
\label{sec:embed}

It is clear that the infinitesimal distance 
\begin{equation}
ds^2=2u(du^2+dv^2)
\end{equation}
does not uniquely determine a manifold. This
then gives rise to the question of just what sort of surfaces can this 
infinitesimal distance represent.  A particular choice of such a surface 
would determine the range of 
variation of the parameters $u$, $v$ which in turn enables the solution of the 
geodesic equations in the case of classical mechanics and the quantum mechanics
of a point particle. It is known that any two-dimensional Riemannian space can 
be embedded in a three-dimensional Euclidean space of indefinite or definite 
signature. In this section we look at a number of natural embeddings and 
discuss their associated geodesics and quantum mechanics. The infinitesimal 
distance that we are dealing with can be embedded in three-dimensional 
Euclidean space $E_3$ via the formulas
\begin{equation}
X=\sqrt{2u}\cos v\,, \qquad Y=\sqrt{2u}\sin v\,,
\end{equation}
\begin{equation}
Z = \frac{\sqrt 2}3\left(F\left(\varphi,\frac1{\sqrt 2}\right)
                           + \sqrt{4u^3-u} \right)\,,
\end{equation}
where $u\geq\frac12$, $v_0 \leq v\leq 2\pi + v_0$, 
$\sin\varphi=\sqrt{2u+1}$ and 
$F(\varphi,k)$ is an elliptic
integral of the first kind.  This embedding gives the infinitesimal
distance
\begin{equation}
dX^2+dY^2+dZ^2=2u(du^2+dv^2)\,.
\end{equation}
To do quantum mechanics on this surface let us first look for 
separable solutions to the free Schr\"odinger equation. 
A typical solution has already been found in the previous section, viz.
\begin{equation}
\Psi=\sqrt{u-\frac{m^2}{4E}}
      \ C_{\frac13}\left(\frac23\sqrt{4E}\left(u-\frac{m^2}{4E}\right)^{3/2}\right)e^{imv}
\end{equation}
where $m$ is an integer.  
As $u\geq\frac12$ and we see that $u=\frac12$ is not
a singular point of the separable equation in $u$, we can impose a 
condition of the form
\begin{equation}
a\Psi\left(\frac12,v\right)+b\Psi_u\left(\frac12,v\right)=0
\end{equation}
together with the periodic boundary condition $\Psi (u,v)=\Psi (u,v+2\pi )$,
which is already satisfied.  If we take $a=1$, $b=0$ then 
$E\geq 0$, otherwise there 
is no solution satisfying the boundary condition at $u=\frac12$.  If $E\geq 0$ 
then we can find a suitably behaved solution that vanishes as 
$u\rightarrow \infty $ and satisfies the boundary condition at $u=\frac12$,
viz.
\begin{equation}
\Psi = (UU')^{1/2}\left( J_{\frac13}(U)J_{-\frac13}(U')
          - J_{\frac13}(U')J_{-\frac13}(U)\right)
\end{equation}
where 
$U=\frac23\sqrt{4E}\left(u-\frac{m^2}{4E}\right)^{3/2}$ and
$U'=\frac23\sqrt{4E}\left(\frac12-\frac{m^2}{4E}\right)^{3/2}$.
These solutions are the analogue of the 
scattering states on this manifold subject to the boundary condition we have 
adopted.

An interesting embedding in pseudo-Euclidean space is given by
\begin{eqnarray}
X &=& \sqrt{2u}v  \nonumber \\
Y &=& \sqrt u\left(\frac45u^2 - v^2 + \frac12\right)  \nonumber \\
T &=& \sqrt u\left(\frac45u^2 - v^2 - \frac12\right) 
\end{eqnarray}
for which $dX^2+dY^2-dT^2=2u(du^2+dv^2)$.  
In this case the variables vary over 
the ranges $-\infty <v<\infty$, $0\leq u<\infty$.  
We could indeed do an 
analysis of the free Schr\"odinger equation on this surface and come to a similar
conclusion if we imposed the condition that the wave function is zero at 
$u=0$.  However if we consider the first potential (\ref{potential1}) and choose 
$b_1=-\beta ^2$, $b_3=\frac14\left(\frac14-\gamma ^2\right)$ for real $\beta $ and 
$\gamma \geq 0$ and if we write the solutions to Schr\"odinger's equation in the form 
$\Psi  =U(u)V(v)$ then two independent solutions of the separation equation 
satisfied by $V$ can be taken as
\begin{equation}
V_\pm =\exp\left(-\frac12\beta v^2\right) v^{\pm\gamma+\frac12}
         {}_1F_1\left(\frac12(1\pm \gamma )- \frac\mu\beta,1\pm \gamma ,\beta v^2\right)\,.
\end{equation}
If we wish to interpret these solutions as being associated with an angle
variable which varies in the range 
$0<v_0 \leq v\leq v_0 +2\pi $ 
and then we would require the periodic boundary conditions 
\begin{eqnarray}
V(v_0) &=& V(v_0+2\pi )  \nonumber \\
V'(v_0) &=& V'(v_0+2\pi )\,.
\end{eqnarray}
The possibility of imposing these boundary conditions depends on whether $v=0$ 
occurs inside the domain of $v$.  If it does not then the spectrum is 
determined from the condition
\begin{equation}
\left.W\left[V_+(x)-V_+(x+2\pi ),V_-(x)-V_-(x+2\pi )\right]\right| _{x=v_0}=0\,.
\end{equation}
If $v=0$ is included then the same conditions no longer work as this is a 
regular singular point of the equation.  Indeed if $v_0=0$ 
and we assume 
$\gamma>\frac12$,
then we choose the solution $V_+$ and impose the condition 
\begin{equation}
V_+(2\pi )=0
\end{equation}
as $V_+(0)$ is already zero.  The quantisation condition is then determined by
\begin{equation}
{}_1F_1\left(\frac12(1+\gamma)-\frac\mu\beta,1+\gamma,4\beta\pi^2\right)=0\,.
\end{equation}
For the $u$ separation equation the range of variation of the variable 
$u>\frac12$
is clear and $u=\frac12$ is not a singularity of the the 
separation equation.  We can accordingly take typical solutions to be 
\begin{equation}
U_\pm (u)=a_1D_\nu\left(2\sqrt\beta\left(u-\frac E{2\beta^2}\right)\right)
      + a_2D_\nu\left(-2\sqrt\beta\left(u-\frac E{2\beta ^2}\right)\right)
\end{equation}
where $\nu =\frac1{4\beta}\left(\frac{E^2}{\beta^2}+4b_2-\mu\right)-\frac12$.  To 
obtain a solution that vanishes as $u\rightarrow \infty $ requires that 
$a_2=0$. The remaining boundary condition becomes
\begin{equation}
D_\nu\left(2\sqrt\beta\left(\frac12-\frac E{2\beta^2}\right)\right)=0\,.
\end{equation}
This condition determines the nature of the discrete spectrum.  For large 
eigenvalues the discrete spectrum is given by
\begin{equation}
E \cong  -2\sqrt{\beta ^3n}
\end{equation}
for suitable large integer $n$.

If we consider the second potential
(\ref{potential2}) then putting $a_3=-\alpha ^2$ the equation for
$V(v)$ has solutions of the form
\begin{eqnarray}
V &=& d_1D_\nu \left(2\sqrt\alpha \left(v-\frac{a_2}{2\alpha ^2}\right)\right)
      + d_2D_\nu\left(-2\sqrt\alpha \left(v-\frac{a_2}{2\alpha ^2}\right)\right) 
         \nonumber \\
 &=& d_1V_++d_2V_-
\end{eqnarray}
where  
\begin{equation}
\nu = \frac1{4\alpha}\left(4\mu +\frac{a^2_2}{\alpha^2}\right)-\frac12\,.
\end{equation}
As there are no singularities in $v$ in this equation and we can require that 
$w_0 \leq v\leq w_0 +2\pi$ 
\begin{eqnarray}
V(w_0) &=& V(w_0+2\pi )  \nonumber \\
V'(w_0) &=& V'(w_0+2\pi )
\end{eqnarray}
which is equivalent to 
\begin{equation}
W\left[V_+(x)-V_+(x+2\pi ),V_-(x)-V_-(x+2\pi )\right]\left| _{x=w_0 }=0\right.\,.
\end{equation}
The solutions for the function $U(u)$ that are well behaved for large $u$ are
\begin{equation}
U(u)=D_\rho\left(2\sqrt\alpha\left(u-\frac E{2\beta^2}\right)\right)
\end{equation}
where $\rho =\frac1{4\beta}\left(4a_1-4\mu+\frac{E^2}{\alpha^2}\right)-\frac12$.

\section{Conclusion}

In this article we have examined one of the four spaces of revolution
listed by Koenigs \cite{Koenigs}.  For the space that we have considered, it has been
shown that there are three potentials that can be added to give
superintegrable Hamiltonian systems of the type we seek.  In each of these
cases we have exhibited the various inequivalent ways in which a
separation of variables can be achieved for both the classical and quantum
equations that result.  This is equivalent to determining the various
inequivalent ways in which a Hamiltonian can be written in Liouville form
(\ref{Hbar}) for suitable separable coordinates $\alpha,\beta$. 
In particular we note
that each of the three superintegrable systems we have examined are such
that when we write out the classical equation $H=E$ and factor out the
denominator we recover a variant of a superintegrable system corresponding
to flat space \cite{FMSUW}.
This is an example of what is called coupling-constant
metamorphosis \cite{HGDR84}.  It has been proven in \cite{TTW} that all of the
superintegrable systems in the plane are such that the bound states
energies can be calculated algebraically.  In all cases the Hamiltonian lies
in the enveloping algebra of $sl(3,\mathbbm R)$.  We conclude that analogous
statements apply to the superintegrable systems that we have found.

\section*{Acknowledgements}
One of the authors (P.~W.)
thanks the Department of Mathematics of the University of Waikato for its
hospitality during his visit.
The research of P.~W.\ was partly
supported by research grants from NSERC of Canada and FCAR du Quebec and
J.~K.\ was supported by the New Zealand Marsden Fund.

\end{document}